\documentclass[useAMS,usenatbib]{mnras}

\DeclareSymbolFont{cmletters}{OML}{cmm}{m}{it}
\DeclareMathSymbol{v}{\mathalpha}{cmletters}{"76}

\usepackage{tabularx,ragged2e,booktabs,caption}
\usepackage{amsmath}
\usepackage{amssymb}
\usepackage{epsfig}
\usepackage{graphicx}
\usepackage{ifthen}
\usepackage{latexsym}
\usepackage{rotating}
\usepackage{times,epsf}
\usepackage{txfonts}
\usepackage{varioref}
\usepackage{verbatim}
\usepackage{array}
\usepackage{url}
\usepackage{color}
\usepackage[dvipsnames]{xcolor}
\usepackage[T1]{fontenc}
\usepackage{epstopdf}
\usepackage{ulem}

\newcommand{\be}{\begin{equation}}
\newcommand{\ee}{\end{equation}}
\newcommand{\bea}{\begin{eqnarray}}
\newcommand{\eea}{\end{eqnarray}}

\newcommand{\Medd}{\dot M_{\rm Edd}}
\newcommand{\Ledd}{L_{\rm Edd}}
\newcommand{\msun}{M_{\odot}}
\newcommand{\Msun}{M_{\odot}}
\newcommand{\mbh}{M_{\rm BH}}
\newcommand{\rg}{R_{\rm G}}

\title[Magnetic flux
  stabilizing thin accretion disks]{Magnetic flux
  stabilizing thin accretion disks}
\author[A. S\k{a}dowski]
       {Aleksander S\k{a}dowski$^{1,2}$\thanks{E-mail: asadowsk@mit.edu (AS)} \\
        $^1$ MIT Kavli Institute for Astrophysics and Space Research,
77 Massachusetts Ave, Cambridge, MA 02139, USA\\
$^2$ Einstein Fellow}

\begin{document}

\maketitle

\label{firstpage}

\begin{abstract}
  We calculate the minimal amount of large-scale poloidal magnetic
  field that has to thread the inner, radiation-over-gas pressure
  dominated region of a thin disk for its thermal stability. Such a
  net field amplifies the magnetization of the saturated turbulent
  state and makes it locally stable. For a $10\msun$ black hole the
  minimal magnetic flux is $10^{24}(\dot M/\Medd)^{20/21}\,\rm G\cdot
  cm^{2}$.  This amount is compared with the amount of uniform
  magnetic flux that can be provided by the companion star --
  estimated to be in the range $10^{22}-10^{24}\,\rm G\cdot cm^2$. If
  accretion rate is large enough, the companion is not able to provide
  the required amount and such a system, if still sub-Eddington, must
  be thermally unstable. The peculiar variability of GRS 1915+105, an
  X-ray binary with the exceptionally high BH mass and near-Eddington
  luminosity, may result from the shortage of large scale poloidal
  field of uniform polarity.
\end{abstract}


\section{Introduction}
\label{s.introduction}

According to the standard model, radiatively efficient, radiation
pressure supported accretion disks are thermally and
viscously unstable \citep{le+instability,ss76,piran-78}. This prediction is in
apparent disagreement with the properties of most black hole (BH)
X-ray binaries. Except for two sources (GRS1915+105 and IGR
J17091-3624, see \cite{belloni+10} and \cite{altamirano+igr}), all
other BH transients stay in their thermal, high/soft states
for days/months without any sign of unstable behavior. The question
arises -- what phenomena does the standard model not account for?

One possiblity for explaining the stability of radiation pressure
dominated thin disks is the presence of a strong magnetic field
which provides additional pressure support and prevents the runaway
heating or cooling that would occur without it. This idea has been
investigated in recent years by \cite{oda+09} and \cite{zheng+11}, and
very recently verified numerically by \cite{sadowski+thin}. Recently, \cite{libegelman-14} have shown that magnetic
fields may help stabilize the disk also through magnetically
driven outflows which decrease the disk temperature and thus help the
disk become more stable at a given accretion rate. The stabilizing effect of strong fields on disk thermal instability
was also discussed by \cite{begelmanpringle-07}.

How to make an accretion disk magnetized enough to prevent thermal
runaway? Magnetorotational instability \citep[MRI,][]{balbushawley-mri} in
an isolated box is
known to saturate at a total to magnetic pressure ratio $\beta=p_{\rm
  tot}/p_{\rm mag}\approx 10$
\citep[e.g.,][]{turner-04,hirose+09}. However, if large scale magnetic
field threads the box, either vertically or radially, the saturated
magnetic field is much stronger. In particular, \cite{xuening+13} have
shown that the presence of a weak net vertical magnetic field characterized
by $\beta_{\rm 0}=1000$ already leads to a saturated state where
magnetic field contributes to roughly half of the total pressure,
which is the rough threshold for the thermal stability
\citep{sadowski+thin}. Consistent results have recently been obtained
by \cite{salvesen+16}.

In this paper we investigate how much magnetic flux contained in large
scale vertical field is needed to stabilize a geometrically
thin, radiatively efficient
accretion flows with radiation pressure dominating over thermal pressure. We compare this quantity with rough estimates of the
magnetic flux that can be provided by the companion stars, and with the
amount of flux required for the magnetically arrested (MAD) state.

Our work is organized as follows. In Section~\ref{s.flux.stablization}
we calculate the magnetic flux required for disk stabilization. The
flux required for the magnetically arrested state is calculated for comparison in
Section~\ref{s.flux.MAD}. In Section~\ref{s.flux.star} we estimate how
much uniform magnetic field a companion star can provide. The discussion is given in
Section~\ref{s.discussion} and our work is summarized in Section~\ref{s.summary}.

\section{Magnetic flux required for disk stabilization}
\label{s.flux.stablization}

In this Section we estimate the minimal amount of poloidal vertical
flux required for thermal stability of an accretion disk. We assume
that the disk is locally stable if at least 50\% of the pressure
is provided by the magnetic field \citep{sadowski+thin}. Such a highly
magnetized state is obtained when disk is threaded by net vertical
magnetic field satisfying $\beta_{\rm 0}\lesssim 1000$, i.e., the
pressure of the large scale component equals to at least one
part in thousand of the sum of radiation and thermal pressures in the
equatorial plane \citep{xuening+13, salvesen+16}. The minimal amount of magnetic flux required for
stability of the entire disk is obtained by integrating the net
vertical flux over the whole otherwise unstable region, i.e., where
radiation pressure exceeds gas pressure. Doing so we also assume that
no magnetic flux has accumulated at the BH.

In the standard $\alpha$-disk model \citep{ss73} the vertically
integrated total pressure $P$ at radius $R$ is determined solely by the angular momentum
conservation,
\be
2 \pi R^2 \alpha P_{\rm tot} = \dot M\left(\sqrt{G\mbh R}-\sqrt{G\mbh R_{\rm in}}\right),
\ee
where $\dot M$ is the accretion rate, $\mbh$ is the mass of the BH,
$G$ is the gravitational constant, $\alpha$ is the disk viscosity
parameter, and $R_{\rm in}=6\rg=6GM/c^2$ is
the location of the inner edge of the disk\footnote{In this work we
  ignore, for simplicity, the BH rotation.}. In cgs units it equals,
\be
\label{e.P}
P_{\rm tot}=8\times 10^{23}
\left(\frac{\dot M}{\Medd}\right)
\left(\frac{R}{R_{\rm G}}\right)^{-3/2}
\left(\frac{\alpha}{0.1}\right)^{-1}
J
\,\rm erg/cm^2,
\ee
where $\Medd=2.48 \times 10^{18}\mbh /M_{\odot}  \,\rm g/s$ is the
Eddington accretion rate (which, according to this definition,
corresponds to a thin disk emitting the Eddington luminosity), and $J=1-\sqrt{R_{\rm in}/R}$.

The strength of the net vertical field that is required to provide
highly magnetized saturated turbulent state is given with respect to
the equatorial plane pressure, not the vertically integrated one. This
may be estimated knowing the disk half-thickness $H$ which for the radiation
pressure dominated regime of the standard thin disk solution equals
\citep{shapiro+book}, 
\be
\label{e.h}
H=2\times 10^{7}
\left(\frac{M_{\rm BH}}{10 M_\odot}\right)
\left(\frac{\dot M}{\Medd}\right)
J
\,\rm cm.
\ee
The equatorial plane total pressure $p_{\rm tot}$ now equals $p_{\rm
  tot}=P_{\rm tot}/2H$.

According to our assumptions, half of the total pressure comes from
the magnetic field. The remaining amount is the sum of the radiation
and thermal pressure. The strength of the net
vertial field, $B^z_{\rm 0}$, that was needed to enhance the
magnetization of the saturated state, is,
\bea
\label{e.Bz0formula}
B^z_{\rm 0}&=&\sqrt{8\pi \frac{p_{\rm tot}}{2\beta_{\rm 0}}}=\\\nonumber
&=& 2\times 10^{7}
\left(\frac{M_{\rm BH}}{10 M_\odot}\right)^{-1/2}
\left(\frac{R}{R_{\rm G}}\right)^{-3/4}
\left(\frac{\beta_{\rm 0}}{1000}\right)^{-1/2} 
\left(\frac{\alpha}{0.1}\right)^{-1/2}
\,\rm G.
\eea

The total required flux is obtained by integrating $B^z_{\rm 0}$ over
otherwise unstable region. The radiation pressure dominates over gas
pressure in the inner region up to a critical
radius $R_{\rm max}$ \citep{shapiro+book},
\be
\label{e.Rmax}
R_{\rm max}/R_{\rm G}=9\times 10^{2}
\left(\frac{M_{\rm BH}}{10 M_\odot}\right)^{2/21}
\left(\frac{\dot M}{\Medd}\right)^{16/21}
\left(\frac{\alpha}{0.1}\right)^{2/21}.
\ee
Performing the integral one obtains\footnote{$1\,{\rm G\cdot
  cm^2}=1\,{\rm Maxwell\,(Mx)}$.}
\be
\label{e.Phi0}
\Phi_{\rm 0}=1\times 10^{24}
\left(\frac{M_{\rm BH}}{10 M_\odot}\right)^{34/21}
\left(\frac{\dot M}{\Medd}\right)^{20/21}
\left(\frac{\beta_{\rm 0}}{1000}\right)^{-1/2} 
\left(\frac{\alpha}{0.1}\right)^{-8/21}
\,\rm G\cdot cm^2.
\ee

When calculating the minimal amount of large scale magnetic flux required
for stabilization of the disk we have implicitly assumed that the
accretion flow, which at the accretion rates of interest is
geometrically thin, was able to advect this net magnetic field inward into
the inner region. Whether or not the large-scale field can be
advected depends on the balance between the
advection and diffusion of the field. Standard geometrically thin
disks drag vertical magnetic field inefficiently
\citep{lubow+94,ghosh+97}, and therefore 
are unlikely to drag significant amount of magnetic field on to
the BH. If magnetic field accumulates there it 
exerts significant outward pressure on the accretion flow (see 
the discussion of the magnetically arrested state in 
Section~\ref{s.flux.MAD}). The net magnetic field that we require for stabilization
does not need to accumulate on the BH --  it is enough
if it threads the disk itself and its pressure is a factor
$\beta_{\rm 0}$ lower that the total pressure. The radial gradient of
such net-field  pressure 
is negligible in thin accretion disks when compared with
gravitational and centrifugal forces. In other words, it is only the
magnetic tension that has to be overcome, not the radial gradient of
pressure (as in the MAD state). Whether or not sub-Eddington disk are
able to drag even such a small amount of field inward, is still
debated \citep[see
e.g.,][]{guletogilvie-12,guletogilvie-13,avara+15}. Our work bases on
the assumption that it is possible.

\subsection{Application to BH binaries}
\label{s.application}

Equation~\ref{e.Phi0} gives the minimal amount of magnetic flux of
uniform polarity that
has to be provided to thermally stabilize radiativelly efficient,
radiation-over-gas pressure dominated thin disks. This amount depends
on the BH mass ($M_{\rm BH}$) and the accretion rate ($\dot M$), as
well as
two other parameters ($\beta_{\rm 0}$ and $\alpha$) that result from
non-linear evolution of MRI and are likely to have rather weak or no dependence
at all on the former two. Therefore, the amount of required magnetic
flux scales mostly with the BH mass and the accretion rate.

There are ${\sim}20$ low-mass X-ray binaries with existing
dynamical estimates of the compact object mass indicating towards BHs
\citep[e.g.,][]{ozel+10}. The masses of the transient objects range
from ${\sim} 5$ to ${\sim} 12\msun$. Most such systems undergo
transitions from the quiescent states to outbursts when they reach
significant fractions of the Eddington luminosity \citep{dunn+10} and
enter the radiation-over-gas pressure dominated, presumably unstable,
regime. 

To estimate how much magnetic flux of uniform polarity is
required for stabilization of each source one needs some measure of
the accretion rate. For this purpose we take the
luminosity at outburst maximum as given in Table~2 of
\cite{steiner+jets}\footnote{Because the thin disk geometry may not
  apply to magnetic pressure dominated disks, we adopt their $L_{\rm Peak}$ that was obtained assuming
  isotropic emission.}. In Table~\ref{t.fluxes} we show the masses,
accretion rate estimates and the required fluxes (Eq.~\ref{e.Phi0},
obtained assuming the fiducial values of $\beta_{\rm 0}$ and $\alpha$)
 for the five BH X-ray
binaries with well established BH masses and existing estimates for
the peak luminosities (compiled from \cite{steiner+jets} and \cite{fragos+15}). The amount of magnetic flux required for
their stabilization ranges from $6.8\times 10^{22}$ to $5.7\times
10^{23}\,\rm G\cdot cm^2$. The source that requires by far most of
uniform large scale magnetic flux (almost
three times more than the second one) is GRS 1915+105. This
particularly high number results from the largest BH mass and
accretion rate which
determine the physical size of the radiation-over-gas dominated region
that has to be stabilized by providing external large scale net
vertical field. GRS 1915+105 is at the same time the only source with
well established parameters that shows long duration, very rapid and short timescale
variability near the Eddington luminosity\citep[e.g.,][]{belloni+10} that results, presumably, from
thermal instability of its accretion disk. One should note, however,
that the exceptionally high mass and luminosity of GRS 1915+105 are
not the only properties making it stand out. At the same time it has
the longest orbital period, the largest accretion disk, a giant
companion star \citep{greiner+01} and presumably also the
largest BH spin \citep{mcclintock+06}.

\begin{table}
\begin{center}
\caption{Magnetic fluxes required to stabilize particular BH X-ray binaries}
\label{t.fluxes}
\begin{tabular}{lccc}
\hline
Name & $M_{\rm BH}/\msun$ & $L_{\rm Peak}/\Ledd$ & $\Phi_{\rm 0}\,[\rm G\cdot cm^2]$\\

\hline
GRS 1915+105 & $12.4$ & $1.0$ & $5.7\times 10^{23}$
\\
XTE J1550-564 & $9.1$ & $0.53$ & $1.9\times 10^{23}$
\\
GRS 1124-683 & $7.0$ & $0.61$ & $1.4\times 10^{23}$
\\
A0620-00 & $6.6$ & $0.47$ & $1.0\times 10^{23}$ \\
GRO J1655-40 & $6.3$ & $0.34$ & $6.8\times 10^{22}$ \\
\hline
\multicolumn{4}{l}{Masses and luminosities compiled from \cite{steiner+jets},}\\
 \multicolumn{4}{l}{\cite{reid+14} and \cite{fragos+15}.}
\end{tabular}
\end{center}
\end{table}

\section{Magnetic flux required for the Magnetically Arrested State}
\label{s.flux.MAD}

Magnetic flux that is advected across the inner edge of the disk 
accumulates on the BH. If advection is efficient the 
accumulated magnetic field exerts radial pressure large enough to
dynamically affect the infalling gas. When this outward magnetic
pressure roughly balances tha radial gravitational force, the disk
enters the magnetically arrested state (MAD), where accretion is
possible only because the interchange instability allows the gas
to penetrate the accumulated field by breaking into clumps or
filaments \citep{narayan+mad,sasha+madjets}.

The amount of magnetic flux accumulated at the BH which results in the MAD
state extending up to a given radius was estimated by
\cite{narayan+mad} (their Eq.~2). Taking the horizon radius as
the limit of the extent of the MAD regime, one obtains the minimal
amount of flux required to provide the saturated magnetic field at the
BH \citep[see also][]{yuannarayan-14},
\be
\label{e.PhiMAD}
\Phi_{\rm MAD}=1\times 10^{23}
\left(\frac{M_{\rm BH}}{10 M_\odot}\right)^{3/2}
\left(\frac{\dot M}{\Medd}\right)^{1/2}
\left(\frac{\epsilon}{10^{-2}}\right)^{-1/2} \,\rm G\cdot cm^2,
\ee
where $\epsilon$ is the ratio of the gas radial velocity to the
free-fall velocity. If only such an amount is advected on the BH, the
innermost part of the flow will be in the MAD state, and the BH
itself, if rotating, will efficiently generate relativistic magnetic
jets \citep{bz}.

\section{Magnetic flux provided by the companion star}
\label{s.flux.star}

The companion stars in most BH X-ray binaries overflow their Roche lobes
and transfer gas to the compact object. The expelled matter forms an
accretion disk and gradually approaches the black hole, sometimes in a
rather violent way due to the ionization instability modulating the
flow \citep{lasota+instability}. The gas drawn from the stellar surface brings
magnetic field with it which triggers turbulence in the accretion
disk. One may expect, that such magnetic field will have some
coherence which will determine the amount of poloidal magnetic flux of
uniform polarity
available in the accretion disk. Below we very roughly estimate that
amount.

It is reasonable to expect that the companion stars in low mass X-ray
binaries are tidally locked to the rotation of the binary system. This
fact has significant consequences. Firstly, the gas that overflows the
Roche lobe comes from exactly the same substellar spot on the companion star
surface. Therefore, the magnetic field advected with the gas towards
the compact object will simply reflect the magnetic field in the
surface layers of the star, and will not be affected by sweeping
through the stellar
surface due to rotation. Secondly, tidal locking modifies the
rotational period of the companion star, which is one of the major
factors determining the efficiency of
magnetic field generation in the stellar interior.

\begin{figure}
\centering
 \includegraphics[width=1.\columnwidth]{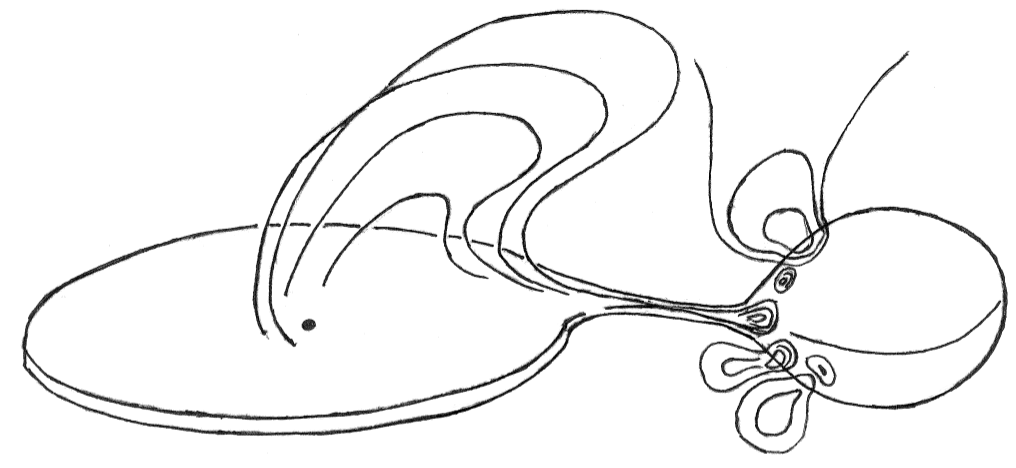}
 \caption{Schematic picture of the advection of magnetic tubes
   from the surface layers of the companion star towards the inner
   region of the accretion disk in an X-ray binary.}
 \label{f.sketch}
\end{figure}

For obvious reasons the magnetic field of the Sun is known best \citep[see
e.g.,][]{schrijverzwaan}. It
exhibits 11 year long activity cycles over which the polarity of the dipolar
component flips. Over that period Solar activity changes as
well. During the activity periods multiple flares, coronal mass
ejections, and sunspots occur
on the surface. These phenomena are related to the emergence of
magnetic field, either in the form of large closed loops or open field
lines, which is generated within the convective envelope of the Sun
through the dynamo process \citep{parker-55}. Emerging magnetic field probes the magnetic
field properties below the stellar surface. In particular, one may
expect that the magnetic field below the surface forms magnetic field
loops containing similar magnetic flux as the loops penetrating the
stellar surface. The magnetic tubes in the surface layers will be
advected with the gas and may dominate the large-scale properties of
the magnetic field in the accretion disk (Fig.~\ref{f.sketch}).

The magnetic field near the active regions in the Sun
can reach and exceed ${\sim} 2000\,\rm G$
\citep{aschwanden-book}. The sizes of  Solar sunspots hardly exceed
$0.001$ of the solar hemisphere area
\citep{harveyzwaan-93}. Multiplying these two numbers (magnetic field
strength by the estimate of its coherence area) one gets the estimate
of the maximal amount of
magnetic flux in an active region (reflecting roughly the amount of magnetic flux in coherent regions below the surface) -- $\Phi_{\odot}\lesssim
10^{23}\,\rm G\cdot cm^2$. \cite{linskyscholler-15} give the range of magnetic fluxes
in individual active regions on the Sun as $10^{20}$ to $10^{22.5}\,\rm
G\cdot cm^2$, roughly consistent with the previous estimate.

Direct measurements of magnetic flux contained in single magnetic
tubes in distant stars is in most cases impossible. Magnetic fields on
stars are implied from the
spatially-unresolved stellar light of nearby stars which provides
information only about the integrated (affected by cancelations of
magnetic field of opposing polarity) magnetic field. Even Zeeman Doppler
imaging \cite[e.g.,][]{donatilandstreet-09} can measure only the net
(again affected by cancellations) magnetic field in active regions.

Companion stars in X-ray binaries differ from the Sun in many
aspects \citep[for rather comprehensive list BH X-ray binaries
properties see][]{fragos+15}. They show different masses and 
spectral types. Most are K or M type dwarfs with masses between $0.2$
and $0.9\Msun$. Some (e.g., the companion of GRS 1915+105) are evolved
giants. Still, all of them have one common property -- rapid rotation
resulting from tidal synchronization of the companion star with the
binary.  Almost half of the systems with well established parameters
of both BH and the companion star have orbital (and rotational) periods
below $0.5\,\rm d$ (the Sun rotates at the equator with ${\sim}24\,\rm
d$ period). Only one BH X-ray binary rotates with a longer period than
the Sun (GRS
1915+105, ${\sim}34\,\rm d$), but this rate of rotation is still
exceptional for a giant star. It is unclear to what level and for how long the tidally
synchronized companions retain differential rotation of their interiors.

Like in the Sun, the stellar
magnetic fields are assumed to result from the dynamo activity in
their differentially rotating convective zones.  This
assumption is supported by observations showing that the activity
indeed scales with rotation, in agreement with the dynamo theory
\citep{reiners-12}.
 This relation can be characterized by the Rossby number, $R_{\rm 0}$ -- the ratio
 of the rotational period of the star and its convective turnover
 time \citep[e.g.,][]{stepien-94}. \cite{reiners+09} have shown that the magnitude of the mean
 surface magnetic field of $K$
 and $M$ dwarfs
saturates
 around $3000\rm\,G$ for $R_{\rm 0}<10^{-1.5}$, and scales like $1/R_{\rm 0}$
 for larger values. The stars with saturated magnetic fields strength
 are the fastest rotators (orbital periods $\lesssim \rm days$) -- quite similar to
 most of the BH X-ray binary companions. The Rossby number of the Sun
 is of the order of unity \citep{reiners-12}. Therefore, if the Sun was
 rapidly rotating and Sun-like stars followed the same dependence on
 the Rossby number, its mean magnetic field would be $10$ to $100$ times
 stronger than it is presently. If the size of the active regions
 stayed the same (a conservative assumption), the magnetic flux contained in a single magnetic
 tube would increase by a similar factor.

 We conclude that the amount of the magnetic flux of uniform polarity that the
 companion star provides depends strongly on the star properties, most
 importantly its rotation, but also mass and evolutionary
 history. A rapidly rotating Sun would
 likely provide at most $10^{22}-10^{24}\,\rm G\cdot cm^2$. It is not
 possible to give a comparable estimate for any of the companion
 stars in X-ray binaries due to our lack of understanding of magnetic
 properties of distant stars, especially ones tidally locked and
 significantly affected by evolution. We therefore take the range
 specific for the fast rotating Sun only as the ballpark for what the
 companion stars
 can provide.

\section{Discussion}
\label{s.discussion}

In the previous sections we estimated magnetic fluxes required
for the thermal stability of a thin accretion disk, for the
MAD state, and provided by the companion
stars. All three of them turn out to have the same order of
magnitude for near-Eddington accretion rates and stellar mass BHs. It
is somewhat surprising, especially when comparing the first two with
the magnetic flux provided by the stellar companion which is
determined by the efficiency of stellar dynamo which knows nothing
about the properties of the inner regions of accretion disks.

\begin{figure}
 \includegraphics[width=.95\columnwidth]{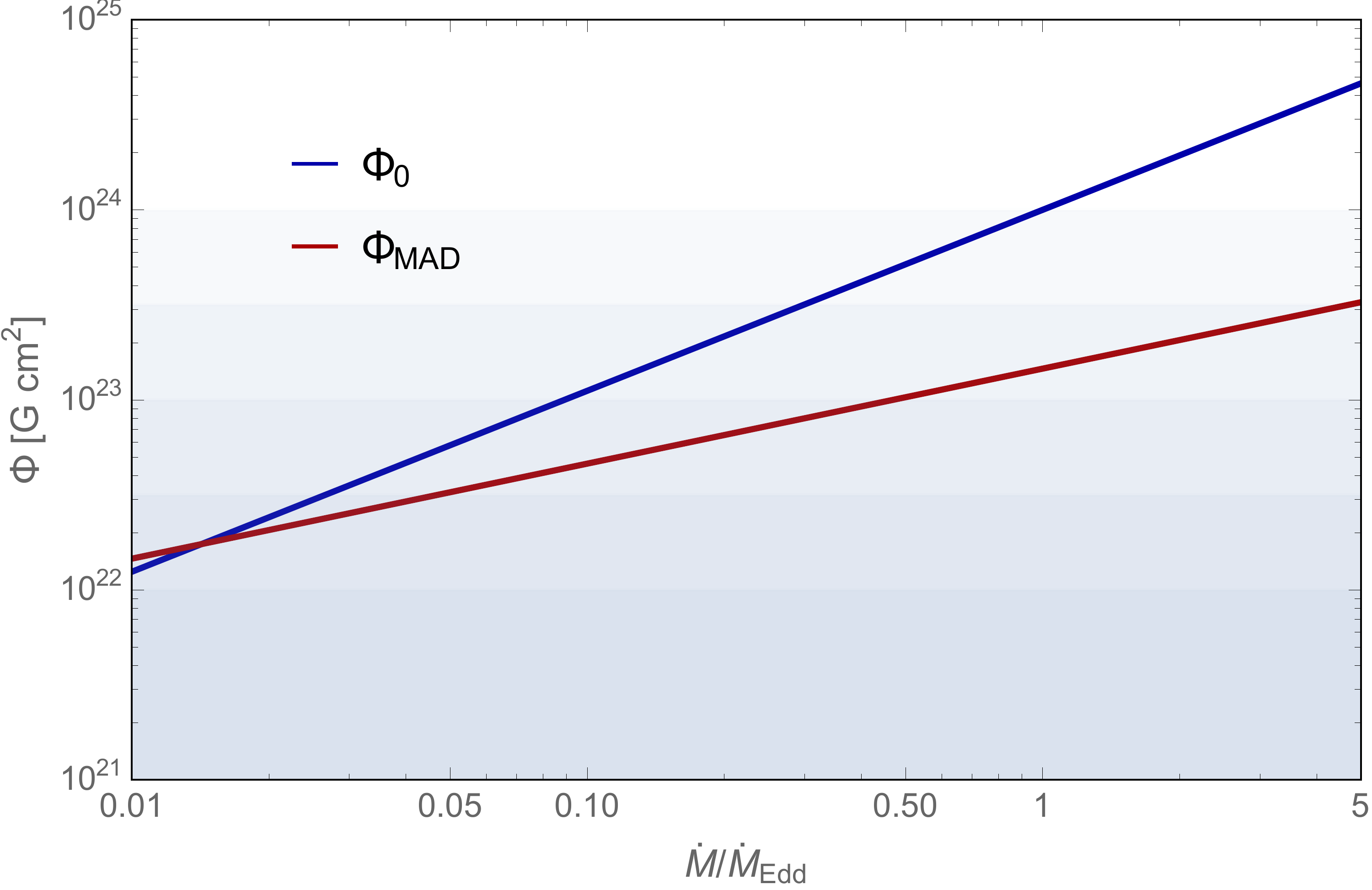}
 \caption{Minimal magnetic flux required for stabilization of a thin disk
   ($\Phi_{\rm 0}$, blue line), as a function of normalized accretion
   rate. Red line denotes the minimal amount of magnetic flux required
   for the magnetically arrested (MAD) state ($\Phi_{\rm MAD}$). These
   fluxes were calculated assuming $M_{\rm BH}=10\msun$, $\beta_{\rm
     0}=1000$, $\alpha=0.1$, and $\epsilon=0.01$. The
   shaded region reflects the rough estimate of the maximal magnetic
   flux provided by the companion star ($10^{22}-10^{24}\,\rm G\cdot
   cm^2$, Section~\ref{s.flux.star}).}
 \label{f.fluxes}
\end{figure}

Figure~\ref{f.fluxes} shows how the fluxes required for stabilization
and for the magnetically arrested state change with accretion rate for the fiducial
parameters ($M_{\rm BH}=10\msun$, $\beta_{\rm
     0}=1000$, $\alpha=0.1$, and $\epsilon=0.01$). Both increase with
   the accretion rate, but the flux required for stabilization
   ($\Phi_{\rm 0}$, blue line) grows faster. At roughly $0.01\Medd$
   they have the same value, ${\sim}10^{22}\,\rm G\cdot
   cm^2$.

The shaded region in the background
   reflects the rough estimate of the amount of the uniform magnetic flux that
   can be provided by the companion star, $10^{22}-10^{24}\,\rm G\cdot
   cm^2$ (Section~\ref{s.flux.star}). It is clear that for a fixed
   value of that quantity, i.e., for a given companion star, there is
   a critical accretion rate above which the flux provided by the
   companion is not enough to stabilize  the unstable inner region of the
   accretion flow. Therefore, each X-ray binary system becomes
   unstable if this critical accretion rate (specific for each system)
   is exceeded. One may
   expect that this critical value is of the order of the Eddington
   accretion rate.

If the companion star is capable of providing plenty of magnetic field
of uniform polarity, then such a critical accretion rate would exceed
the Eddington one and a given system will never become unstable
since super-Eddington accretion flows are stabilized by advection of
heat \citep[see][]{abramowicz+slim}.

Qualitatively similar conclusion applies to the amount of flux
required for the magnetically arrested disk -- even if the advection
of magnetic field allows for the field accumulation on the BH, there
exists a critical, near-Eddington accretion rate above which the MAD state cannot be
sustained. In other words, highly super-Eddington accretion flows
cannot be magnetically arrested and efficiently produce
magnetically-driven, relativistic jets \citep[although they are likely to
generate radiative jets, see][]{sikorawilson-81,narayan+83,sadowski+radjets}.

If only the companion star provides the large scale
poloidal magnetic flux and this critical accretion rate for
stabilization is indeed significant, then one
would expect that X-ray binaries with most massive BHs and accreting
at largest, but sub-Eddington, accretion rates will be most difficult
to stabilize. Out of the best known BH X-ray binaries, GRS 1915+105 is
such an example, with most massive BH and near-Eddington outburst
luminosity, and it is indeed the only one unstable. It has to be
mentioned, however, that its companion star is very evolved and
presumably significantly affected by binary evolution \citep{fragos+15}, and
therefore peculiar within the set of other companion stars. Thus, the
estimate of the magnetic flux available from the companion that we
derived basing on Solar and dwarf star magnetic properties may not be
accurate \citep{stepien-94}.

Thermal instability is not specific to BH accretion
  flows. Similar phenomenon is expected to take place in radiation
  pressure dominated, radiatively efficient disks around neutron
  stars (NSs). Similarly to the BH case, large scale magnetic field may play important
  role in stabilizing them. In the case of NS systems, however, the
  magnetic field of the NS itself may provide extra stabilizing
  effect. In addition, low mass of NSs would suggest small amount of
  magnetic flux required for stabilization, relatively easier to
  provide by the companion star. Out of all NS systems known, only the
Rapid Burster showed (twice in 16 years) light curves that resemble
those of GRS 1915+105 \citep{bagnoli+15} what seems to be consistent
with the picture presented in this Letter.

Several questions, however, arise. We assumed that the magnetic flux
is provided by the companion star, and the magnetic tubes reaching its
surface layers near the substellar point are efficiently
dragged into the inner region of the accretion disk. To provide the
observed stability of outbursts of most X-ray binaries, the duration of
which is determined by the propagation of viscous instability through
the whole accretion disk and is often of the order of months \citep{lasota+instability}, one would
have to make sure that magnetic flux accumulated in the inner region
is not canceled out by a flux of opposite polarity during this
period. That would require either that a single magnetic tube is
accreted for a longer time than the outburst duration, or that the net
magnetic field of the tubes hover for such a time at fixed location in the disk having
established the advection/diffusion equilibrium \citep[analytical
solutions of the steady-state radial distribution of poloidal fields
were obtained and studied by][]{okuzumi+14}. The latter may be
preffered if thin accretion disks are indeed inefficient accretors of
the magnetic field.

One other question is whether it is indeed the magnetic field from the
active regions of the
companion star that dominates the large scale magnetic structures in
the disk. In principle, plasma-related effects may be operating as
well. An example is the Contopoulos battery which can generate
poloidal magnetic flux of uniform polarity as a result of the
Poyinting-Robertson radiative drag
\citep{contopoulosnathanail+15}. Although field generated in
such a way is instantenously insignificant, given enough time, it could
aggregate to provide the amount of magnetic flux relavant in the
context discussed in this work.

\section{Summary}
\label{s.summary}

We have calculated the minimal amount of large-scale poloidal magnetic
flux that has to thread the inner part of a thin, radiation-over-gas
pressure dominated accretion disk to stabilize it against thermal
instability. We have compared that amount with the magnetic flux that
has to accumulate on the BH to magnetically arrest the disk, and with
the maximal magnetic flux of uniform polarity that can be advected
with the gas from the companion star in X-ray binaries. We summarize
our findings below, all of which depend on the assumptions that 
magnetic field can be advected inward and that the magnetic field 
coming from the companion star dominates the
large-scale magnetic properties of the inner accretion disk region.

\begin{enumerate}
\item \textit{Magnetic flux required for stabilization:} -- To
  stabilize the inner region of a thin accretion disk, where
  radiation pressure
  dominates over thermal pressure, one has to provide net poloidal
  flux of the order of $10^{23}\,\rm G\cdot cm^2$ for $\dot
  M=0.1\Medd$ and $10\msun$ BH (Eq.~\ref{e.Phi0}). Such a magnetic
  field, although weak when compared with the local gas and radiation
  pressures, will enhance the magnetization of the saturated state of
  MRI and lead to a magnetic pressure supported, and
  therefore stable, state. This critical
  amount of the large-scale magnetic flux grows with BH mass and the
  normalized accretion rate. This net magnetic field does not have to 
  accumulate on the BH and therefore
  can be relatively easily advected into the inner region.

\item \textit{GRS 1915+105:} -- Out of the BH X-ray binaries with well
  established BH and binary parameters, GRS 1915+105 requires most
  magnetic flux, ${\sim}6\times 10^{23} \,\rm G\cdot cm^2$ to be 
  stabilized due to its large BH mass and luminosity.

\item \textit{Magnetic flux provided by the companion star} -- In a
  tidally locked X-ray binary the gas accreting towards the compact
  object can drag magnetic field from the surface layers of the companion
  star. We estimated the amount of magnetic flux contained in magnetic
  tubes of rapidly rotating stars to be of the order of
  $10^{22}-10^{24}\,\rm G\cdot cm^2$.

\item \textit{Critical accretion rate for stabilization} -- For a given
  system, the amount of magnetic flux required for stabilization above
  some critical, near-Eddington, accretion rate is larger than can be provided by the
  companion star. Such systems are expected to be thermally unstable
  (and GRS 1915+105 may be an example),
  unless their transfer rate exceeds the Eddington rate, in which
  case they are stabilized by the advection, and magnetic contribution
  is no longer required.

\item \textit{Magnetically Arrested Disk} -- To saturate the magnetic
  flux accumulated at the BH, and to enter the MAD state resulting in
  efficient jet production, one has to provide a comparable amount of
  magnetic flux (${\sim} 10^{23}\,\rm G\cdot cm^2$ for $1 \Medd$, Eq.~\ref{e.PhiMAD}). This
  state, however, requires significant accumulation of the magnetic
  field at the BH that exerts outward pressure and therefore requires
  very efficient advection of the magnetic field, which may not be the
  case for thin disks. Even if the advection is effective, when the
  accretion rate exceeds significantly the Eddington rate, the
  companion star cannot provide enough uniform magnetic flux to
  maintain the magnetically arrested state. Therefore, one should not
  expect efficient generation of relativistic jets in super-Eddington accretion flows.

\end{enumerate}

\section{Acknowledgements}

The author thanks Jack Steiner, Jean-Pierre Lasota, Ramesh
Narayan, and John Raymond for helpful comments.
The author acknowledges support
for this work 
by NASA through Einstein Postdoctotral Fellowship number PF4-150126
awarded by the Chandra X-ray Center, which is operated by the
Smithsonian
Astrophysical Observatory for NASA under contract NAS8-03060. 
\bibliographystyle{mn2e}

\end{document}